\documentclass[aps,prc,showpacs,12pt]{revtex4}
 \usepackage{graphics}
 \usepackage{amsmath}
 \usepackage{epsfig}
 \usepackage[english]{babel}

\begin{document}

\title{Theory of the nuclear excitation by electron transition process
near the $K$-edge.\footnote{Accepted for publication in PRA.}}

\author{E.V.~Tkalya}

\email{tkalya@srd.sinp.msu.ru}

\address{Institute of Nuclear Physics, Moscow State University,
Ru-119992, Moscow, Vorob'evy Gory, Russia}

\date{\today}

\begin{abstract}
We propose a model for description of the process of Nuclear
Excitation by Electron Transition (NEET) near the $K$-shell
ionization threshold of an atom. We explain the experimental results
for the $^{197}$Au cross section excitation $\sigma_{N^*}$ obtained
by S.Kishimoto et al. Phys. Rev. C {\bf 74}, 031301(R) (2006) using
synchrotron radiation near the Au $K$-edge. We predict the
behavior of $\sigma_{N^*}$ as a function of the incident photon energy
for nuclei $^{193}$Ir and $^{189}$Os. We reveal that the $^{189}$Os
excitation begins when the energy of incident photons is below the
$K$-shell ionization threshold in Os.
\end{abstract}


\maketitle

\section{Introduction}

Nuclear Excitation by Electron Transition (NEET)
\cite{Morita-73} is a process of nonradiative nuclear excitation by
means of direct energy transfer from the excited atomic shell to the
nucleus via a virtual photon. This process is possible, if within the
atomic shell there exists an electronic transition close in energy
and coinciding in type with the nuclear one.

The modern theory of the NEET was developed in
Refs.~\cite{Tkalya-92-NP,Tkalya-92-JETP,Tkalya-00,Ahmad-00,Harston-01,Tkalya-94-JETP}.
This theory agrees reasonably with experimental results obtained for
nuclei $^{197}$Au \cite{Kishimoto-00}, $^{193}$Ir
\cite{Kishimoto-05}, and $^{189}$Os \cite{Ahmad-00} in recent years.

The NEET process was investigated in the experiments
\cite{Kishimoto-00,Kishimoto-05,Ahmad-00} for the case where the
energy of photons exceeded the $K$-shell electron binding energy
appreciably. In such a situation one could neglect the threshold
effects \cite{Tkalya-92-NP,Tkalya-92-JETP,Tkalya-00} and write the
relative probability for nuclear excitation in atomic transition
$P_{NEET}$ in the following form

\begin{equation}
P_{NEET}=\left(1+\frac{\Gamma_{i}}{\Gamma_{f}}\right)
\frac{E^2_{int}}{(\omega_N-\omega_A)^2+(\Gamma_{i}+\Gamma_{f})^2/4}\,.
\label{eq:Pneet}
\end{equation}
In Eq~(\ref{eq:Pneet}) $\Gamma_{i,f}$ are the widths of the initial
and final electronic states, $\omega_A$ and $\omega_N$ are the energies
of the atomic and nuclear transitions (the adopted system of units is
$\hbar=c=1$), $E^2_{int}$ is averaged over the initial states and
summed over the final ones, the square modulus of the Hamiltonian of
the interaction $H_{int}$ of the electronic hole current
$j^{\mu}_{fi}({\bf{r}})$ and the nuclear current
$J^{\nu}_{fi}({\bf{R}})$ in the second order of the perturbation
theory for the QED

\begin{equation}
H_{int}=\int{}d^3rd^3Rj^{\mu}_{fi}({\bf{r}})
D_{\mu\nu}(\omega_N,{\bf{r}}-{\bf{R}})J^{\nu}_{fi}({\bf{R}}) \,,
\label{eq:Hint}
\end{equation}
where $D_{\mu\nu}(\omega_N,{\bf{r}}-{\bf{R}})$ is the photon
propagator.

The cross section of $^{197}$Au excitation on the isomeric level
$1/2^+$(77.351 keV, 1.91 ns) by photons near the $K$-shell ionization
threshold of gold was measured in work \cite{Kishimoto-06}. It was
found that the NEET events rose up just above the $K$-absorption edge
and the NEET edge of width $14\pm9$ eV existed at $40\pm2$ eV higher
than the $K$-edge. The incident synchrotron radiation beam had a
$3.5\pm 0.1$ eV width.

Theoretical explanation is needed for the results obtained in
Ref.~\cite{Kishimoto-06}. We consider here a model of the process
founded in Ref.~\cite{Kishimoto-06} and describe the main
characteristics of the process. Following the tradition we call this
process ``the NEET near $K$-edge''. However, here we pay attention to
the cross-section of the process of the nuclear excitation as a
result of ionization of the atomic $K$-shell. The NEET as we shall
see below is a special case of the considered process in the
asymptotic limit of ``high'' energies far from the photoionization
threshold.

\section{Model of NEET process near $K$-edge}

\begin{figure}[h]
\includegraphics[angle=-90,width=10cm]{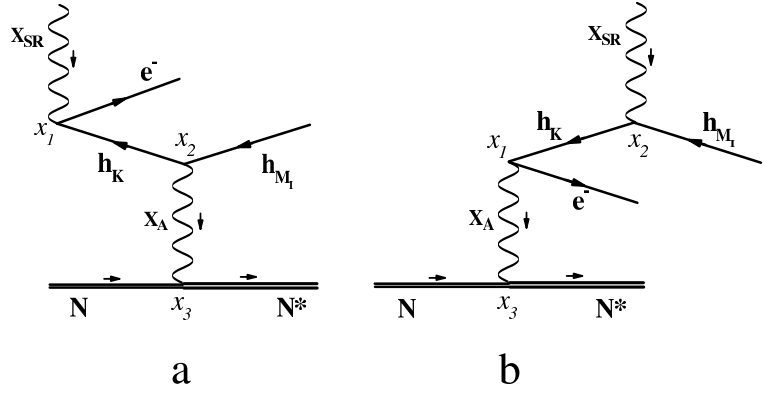}
\caption{Nuclear excitation by ionization of the atomic $K$-shell.
(a) --- direct diagram, (b) --- exchange diagram.}
\label{fig:FD}
\end{figure}

The process of the nuclear excitation as a result of ionization of
the atomic $K$-shell is described by two diagrams shown in
Fig.~\ref{fig:FD}. One electron passes from the $K$-shell to the
continuum. We consider here the case of nuclear excitation by the
atomic $M_I\rightarrow{}K$ transition.


According to the QED rules \cite{Berestetskii-80} the amplitude can
be written as

\begin{equation}
S^{(dir)}_{fi} = -i\int{}d^4x_1d^4x_2d^4x_3
\bar{\psi}_e(x_1)e\gamma^{\mu}A_{\mu}(x_1)G_{h}(x_1,x_2)e\gamma^{\nu}
\psi_{h_{M_I}}(x_2) D_{\nu\rho}(x_2,x_3) J^{\rho}(x_3)
\label{eq:Sdir}
\end{equation}
for the direct diagram (see Fig.~\ref{fig:FD}(a)) and as

\begin{equation}
S^{(ex)}_{fi} = -i\int{}d^4x_1d^4x_2d^4x_3
\bar{\psi}_e(x_1)e\gamma^{\mu} D_{\mu\rho}(x_1,x_3)
G_{h}(x_1,x_2)e\gamma^{\nu}A_{\nu}(x_2) \psi_{h_{M_I}}(x_2)
J^{\rho}(x_3) \label{eq:Sex}
\end{equation}
for the exchange diagram (see Fig.~\ref{fig:FD}(b)). In
Eqs~(\ref{eq:Sdir})-(\ref{eq:Sex}) $e$ is the proton charge,
$\gamma^{\mu}$ are the Dirac matrices. We will use the following
designations for the wave functions of the particles and for the
nuclear current shown in Fig.~\ref{fig:FD}.
$A_{\mu}(x;\omega_{SR})=e^{-i\omega_{SR}t}A_{\mu}({\bf{r}};\omega_{SR})$
is the synchrotron radiation photon with the energy $\omega_{SR}$.
$\bar{\psi}_e(x;E_e)=e^{i{E_e}t}\bar{\psi}_e({\bf{r}};E_e)$ is the
electron with the energy $E_e$ emitted from the $K$-shell to the
continuum. $\psi_{h_{M_I}}(x;E_{M_I}) =
e^{-i(E_{M_I}-i\Gamma_{M_I}/2)t}\psi_{h_{M_I}}({\bf{r}};E_{M_I})$ is
the electron hole at the $M_I$-shell with the binding energy
$E_{M_I}$ and the width $\Gamma_{M_I}$. The photon propagator
$D_{\mu\nu}(x,x_3;\omega_N) =
i\langle0|\hat{T}A^*_{\mu}(x)A_{\nu}(x_3)|0\rangle$ is calculated
according to the relation
$$
D_{\mu\nu}(x,x_3;\omega_N) = \int\frac{d\omega}{2\pi}
e^{i\omega_N(t-t_3)} D_{\mu\nu}({\bf{r}}-{\bf{R}};\omega_N) \,.
$$
The nuclear current is $J^{\rho}({\bf{R}},t_3)=
e\Psi^+_{N^*}({\bf{R}},t_3)\hat{J}^{\rho}\Psi_N({\bf{R}},t_3)$, where
the nuclear wave functions are $\Psi_N({\bf{R}},t_3) = e^{-iE_Nt_3}
\Psi_N({\bf{R}})$ for the ground state ($E_{N}=0$ usually) and
$\Psi^+_{N^*}({\bf{R}},t_3)= e^{i(E_{N^*}+i\Gamma_{N^*}/2)t_3}
\Psi^+_{N^*}({\bf{R}})$ for the isomeric state with the energy
$E_{N^*}$ and the width $\Gamma_{N^*}$, and $\hat{J}^{\rho}$ is the
operator of the nuclear electromagnetic transition. The propagator of
the hole $G_{h}(x_1,x_2)=
-i\langle0|\hat{T}\psi_{h}(x_1)\bar{\psi}_{h}(x_2)|0\rangle$ in our
model is the $K$-shell hole propagator
$$
G_{h_K}(x_1,x_2;E_K) =
-\psi_{h_{K}}({\bf{r}}_1)\bar{\psi}_{h_{K}}({\bf{r}}_2)
\int\frac{dE}{2\pi}\frac{e^{iE(t_2-t_1)}}{E-E_K-i\Gamma_K/2}\,,
$$
where $E_K$ is the electron binding energy, $\Gamma_K$ is the total
hole width at the $K$-shell (see in Ref.\cite{Berestetskii-80}).

The model considered here has some peculiarities. Let us look at some
of them more carefully. Let us begin from the NEET process. The
excitation of the nuclei $^{197}$Au, $^{193}$Ir, and $^{189}$Os takes
place in reality in the electron transition from the $M_I$-shell to
the $K$-shell, which occurs when a photon is absorbed, and electron
in the $K$-shell is excited into the continuum, leaving the $K$-hole
behind. The electrons in the atomic shell interact with each other
and generate a self-consistent field providing that many-particle
effects became very important for most of the processes inside the
atomic shell. That is why one should take into account the ejected
electron-hole interaction, the pair electron-electron interactions,
etc. for correct description of the NEET process. Nevertheless, in
our model the nucleus is excited in a single-particle hole
transition, as it is shown on the diagrams in Fig.~\ref{fig:FD}.
Correspondingly we use the single-particle approximation for the
calculation of the atomic matrix elements. One can do that, because
numerous computations show that this approximation works well for the
electronic transitions between the inner atomic shells, if one
estimates of the $X$-rays intensities, as well as for the electron
transitions from the inner atomic shells to the continuum, if one
estimates of the internal electronic conversion probabilities. That
is why one can use this approximation to the numerical evaluation of
the ``electronic part'' of the NEET.

We use the experimental values for the atomic shell widths. Therefore
$\Gamma_K$ in the propagators does not involve the NEET partial
width. On the other hand the NEET probabilities are very small, and
they do not change practically the $K$-shell widths. For another
thing we use the widths $\Gamma_{M_I}$ for the final hole states on
the diagrams in Fig.~\ref{fig:FD}. This method is not typical for the
QED. As it was shown in Ref.\cite{Tkalya-92-JETP}, the use in the
wave functions of the real widths of the vacancies enables us to take
into account the further decay of the electron hole into an upper
atomic level without having to consider higher-order diagrams.

Another important approximation concerns the hole propagator.
Strictly speaking all the atomic shells, which have the electrons,
contribute to the hole propagator. However the energy conservation
law leaves the $K$-shell only in the hole propagator in the direct
diagram in our case, because only one atomic transition (the
$M_I\rightarrow{}K$ transition) satisfies the condition
$\omega_A\simeq{}\omega_N$. The hole propagator in the exchange
diagram may contain of the contributions of all atomic shells.
However this propagator does not have poles in the discrete part of
the spectrum, hence the contribution of the exchange diagram is
negligible. As for the continuum spectrum there is no any
contribution of these hole states to the hole propagator in
Fig.~\ref{fig:FD}. A photon is absorbed by an electron in the atomic
shell and a hole is arisen. The atomic inner-hole state is
highly-excited and short-lived discrete state. In the general case
the wave function of the time-dependent state can be expressed as a
decomposition into a sum, and an integral over the complete set of
eigenfunctions $\psi_{E_n}({\bf{}x},t)$ with $E_n$ belonging to the
discrete spectrum ($E_n < 0$), and $\psi_{\varepsilon}({\bf{}x},t)$
with $\varepsilon$ belonging to the continuous spectrum. Thus the
propagator of the atomic inner-hole generally should have two parts.
But there are no electrons in the continuum in our case. The
continuum is occupied by holes (i.e. by the electron vacancies).
These electron vacancies are fermions, hence the hole creation is
impossible in these states, and the transitions to these intermediate
states are forbidden too. So, these states do not give any
contribution to the hole propagator at all. It becomes evident, if
one considers the following process (which is identical to the one in
Fig.~\ref{fig:FD}): the hole in the continuum absorbs a photon and
passes to the $K$ shell firstly; after that the nucleus is excited in
the hole transition $K\rightarrow{}M_I$. Only the atomic states
occupied by electrons are the ``acting'' intermediate states here,
similarly to the process in Fig.~\ref{fig:FD}. That is why we
actually can neglect the continuum part of the energy spectrum in the
hole propagator in Eqs~(\ref{eq:Sdir})--(\ref{eq:Sex}). As it follows
from the comparison of the results of
\cite{Tkalya-92-NP,Tkalya-92-JETP,Tkalya-00,Ahmad-00,Harston-01}, the
single-particle approximation describes well the NEET process with
the simplest form of the propagator, and the technique of Feynman
diagrams gives the same formula for the NEET probability, that
follows from solution of the equations for time-dependent amplitudes
of the total ``the hole + the nucleus'' wave function
\cite{Ahmad-00}.

The cross section of the process shown in Fig.~\ref{fig:FD} is
calculated from the following formula \cite{Berestetskii-80}
$$
\sigma_{N^*}=\int\frac{d^3p_e}{(2\pi)^3}
\frac{\sum'|S^{(dir)}_{fi}+S^{(ex)}_{fi}|^2}{T} \,,
$$
where $\sum'$ means averaging over the initial states and summation
over the final ones. The process progresses in the time interval $T$.
We do not fix the electron energy $E_e$. That is why we integrate the
cross section over the electron momentum $p_e$.

It is easy to show that after integration in
Eqs~(\ref{eq:Sdir})-(\ref{eq:Sex}) over the times $t_1,t_2,t_3$ and
over the energies of the intermediate states $E,\omega$ the amplitude
$S^{(ex)}_{fi}$ contains only one resonance condition. This condition
corresponds to the energy conservation law of the process. The
amplitude $S^{(dir)}_{fi}$ has an additional resonance condition for
the energy of photons $\omega_{SR}=E_e-E_K$. As a result the exchange
diagram gives a contribution three orders of the magnitude smaller
than the direct diagram. Correspondingly, one can neglect the
exchange diagram Fig.~\ref{fig:FD}(b) near the threshold. The
following expression is obtained for the cross section in this case

\begin{equation}
\sigma_{N^*}=\int\frac{d^3p_e}{(2\pi)^3} \frac{(2\pi)^2}{\Gamma_K}
f(E_e) \frac{1}{2}\frac{1}{2J_N+1} \sum_{\lambda_{SR}}
\sum_{m_e,m_{h_{M_I}}} \sum_{m_N,m_{N^*}} |H_{ion}|^2|H_{int}|^2 \,.
\label{eq:sigma1}
\end{equation}
In Eq~(\ref{eq:sigma1}) we sum up over the photon polarization
$\lambda_{SR}$, the magnetic quantum number of the free electron
$m_e$, the magnetic quantum number of the $M_I$-hole $m_{h_{M_I}}$,
and the magnetic quantum numbers of the nucleus $m_N,m_{N^*}$. The
function $f(E_e)$ is

\begin{equation}
f(E_e)= \frac{1}{\pi}\frac{\Gamma_K/2}
{(\omega_{SR}-E_e+E_K)^2+\Gamma_K^2/4}
\frac{1}{\pi}\frac{(\Gamma_{M_I}+\Gamma_{N^*})/2}
{(\omega_{SR}-E_e+E_{M_I}-\omega_N)^2+(\Gamma_{M_I}+\Gamma_{N^*})^2/4}\,.
\label{eq:f}
\end{equation}

We introduced two new amplitudes $H_{int}$ and $H_{ion}$ in
Eq~(\ref{eq:sigma1}). The amplitude $H_{int}$ describes an
interaction between the electronic hole current
$j^{\nu}_h({\bf{r}}_2)=
e\bar{\psi}_{h_K}({\bf{r}}_2)\gamma^{\nu}\psi_{h_{M_I}}({\bf{r}}_2)$
and the nuclear current
$J^{\rho}({\bf{R}})=e\Psi^+_{N^*}({\bf{R}})\hat{J}^{\rho}\Psi_N({\bf{R}})$
in the NEET process Eq~(\ref{eq:Hint}).

The expression Eq~(\ref{eq:Pneet}) for the probability of nuclear
excitation by the electron $i\rightarrow{}f$ ($M_I\rightarrow{}K$)
transition $P_{NEET}$
can be adapted to the considered case of the nuclear excitation by
the electronic hole $K\rightarrow{}M_I$ transition by the replacement
$\Gamma_{i,f}\rightarrow{}\Gamma_{M_I,K}$. The interaction energy
$E_{int}$ squared is, by definition, as follows

$$
E_{int}^2 = \frac{1}{2j_{h_K}+1}\frac{1}{2J_N+1}
\sum_{m_{h_K},m_{h_{M_I}}} \sum_{m_N,m_{N^*}} |H_{int}|^2 \,.
$$
It has the following form \cite{Tkalya-00,Ahmad-00,Harston-01}

\begin{equation}
E^2_{int} = 4 \pi e^2 \omega_N^{2(L+1)}
\frac{(j_i1/2L0|j_f1/2)^2}{[(2L+1)!!]^2} \left|{\cal
R}_L^{E/M}(\omega_N)\right|^2 B(E/M\,L;J_i\rightarrow J_f) \,,
\label{eq:Eint}
\end{equation}
where $B(E/ML;J_i\rightarrow J_f)$ is the nuclear reduced probability
\cite{Bohr-69}, and ${\cal R}_L^{E/M}$ are the atomic radial matrix
elements of the electric/magnetic ($E/M$) multipolarity $L$
\cite{Tkalya-92-NP,Tkalya-92-JETP,Tkalya-00}, $(j_i1/2L0|j_f1/2)$ is
the Clebsch-Gordan coefficient, $J_{i,f}$ and $j_{i,f}$ are the
angular momentums of the nuclear and the electronic states,
correspondingly. It is obvious that the functions $E_{int}$ and
$P_{NEET}$ do not depend on the energies $\omega_{SR}$ and $E_e$.

It should also be stated that we take $B(E/ML;J_i\rightarrow J_f)$
from the experimental data. We will consider below $M1$ transitions
from the ground state to the first excited state in the nuclei
$^{197}$Au, $^{193}$Ir, and $^{189}$Os. $B(M1)$ for these transitions
are known and can be taken form the tables
\cite{Xiaolong-05,Wu-03,Artna-Cohen-98}. As for the atomic $M1$
radial matrix elements they are calculated according the formula
\cite{Tkalya-92-NP,Tkalya-92-JETP,Tkalya-00}
$$
{\cal R}_1^M(\omega) = (\kappa_i+\kappa_f) \int_0^{\infty}dr r^2
h_1^{(1)}(\omega r) [g_i(r)f_f(r)+f_i(r)g_f(r)]\,,
$$
where $\kappa=(l-j)(2j+1)$, $l$ is the orbital angular momentum,
$h_L^{(1)}$ is spherical Hankel function of the first kind
\cite{Abramowitz-64}, $g(r)$ and $f(r)$ are, respectively, the large
and small components of the electronic wave functions, with the
condition of normalizing $\int_0^{\infty}dr r^2(g^2(r)+f^2(r)) = 1$.
Here these electron wave functions were evaluated by solving the
relativistic Dirac-Fock equations in the self-consistent field of the
electron shell taking into account the finite nuclear size. The
computer program is described in detail in Ref. \cite{Band-79}.

The amplitude $H_{ion}$ corresponds to the process of the $K$-shell
ionization by a photon
$$
H_{ion} = -i\int{}d^3r_1
e\bar{\psi}_e({\bf{r}}_1)\gamma^{\mu}\psi_{h_K}({\bf{r}}_1)
A_{\mu}({\bf{r}}_1;\omega_{SR})\,.
$$

The cross section of the ionization process has the following form
\cite{Berestetskii-80}

\begin{equation}
d\sigma_{ion} = 2\pi\delta(\omega_{SR}-E_e+E_K) \frac{1}{2}
\sum_{\lambda_{SR}} \sum_{m_e,m_{h_K}}
 |H_{ion}|^2 \frac{d^3p_e}{(2\pi)^3}\,,
\label{eq:sigma-ion}
\end{equation}
where $p_e$ is the electron momentum. Let us take into account that
the expression $|H_{ion}|^2 p_e {\cal{E}}_e$
(${\cal{E}}_e=\sqrt{p_e^2+m^2}$, $m$ is the electron mass) depends
little on the nonrelativistic kinetic energy of electron
$E_e=p_e^2/2m$ near the threshold \cite{Berestetskii-80}. As a
consequence we can introduce the following cross section near the
threshold

$$
\sigma_{ion}^{0}=\lim_{p_e\rightarrow 0} \int 2\pi\frac{1}{2}
\sum_{\lambda_{SR}} \sum_{m_e,m_{h_K}}
 |H_{ion}|^2 p_e {\cal{E}}_e \frac{d\Omega_e}{(2\pi)^3},
$$
and we can consider
$\sigma_{ion}^{0}$ as a constant in the energy range around the
threshold.

The width of the $K$-shell plays an important role in the process
considered here. Let us ``spread'' out the delta-function in
Eq~(\ref{eq:sigma-ion}) over the width $\Gamma_K$. Delta-function is
a limit of the $\delta$-shaped Cauchy sequence. Therefore we can
substitute $\delta(\omega_{SR}-E_e+E_K) \rightarrow
(1/2\pi)\Gamma_K/((\omega_{SR}-E_e+E_K)^2+\Gamma_K^2/4)$ in
Eq~(\ref{eq:sigma-ion}). Now if we compare the obtained expression
with Eqs~(\ref{eq:sigma1})-(\ref{eq:f}) we get the following formula
for the nuclear excitation cross section

\begin{equation}
\sigma_{N^*} = \sigma_{ion}^0 E_{int}^2 \frac{2\pi}{\Gamma_K}
\int_0^{\infty}f(E_e)dE_e \,.
\label{eq:sigma2}
\end{equation}

When the incident photon energy $\omega_{SR}$ is near the
$K$-threshold, the function $f(E_e)$ in Eq~(\ref{eq:f}) decreases
quickly when $E_e$ increases in the range 0 --
($\Gamma_K+\Gamma_{M_I}$). That is why one can integrate in the range
$0\leq{}E_e\leq\infty$ in Eq~(\ref{eq:sigma2}) in spite of the fact
that $E_e$ is the nonrelativistic energy of electron. The integral is
calculated analytically:

\begin{equation}
\int_0^{\infty}f(E_e)dE_e =
\frac{F_1+F_2+F_3}{((\omega_N-\omega_A)^2+(\Gamma_K+\Gamma_{M_I})^2/4)
((\omega_N-\omega_A)^2+(\Gamma_K-\Gamma_{M_I})^2/4)} \,,
\label{eq:Int1}
\end{equation}
where
\begin{eqnarray}
\label{eq:F3} F_1&=&\frac{\Gamma_K}{2} \left( (\omega_N-\omega_A)^2 +
\frac{\Gamma_K^2-\Gamma_{M_I}^2}{4} \right) \left( \frac{1}{2} +
\frac{1}{\pi}\arctan{\left(\frac{\omega_{SR}+E_{M_I}-\omega_N}
{\Gamma_{M_I}/2}\right)} \right)\,, \nonumber
\\
F_2&=&\frac{\Gamma_{M_I}}{2} \left( (\omega_N-\omega_A)^2 -
\frac{\Gamma_K^2-\Gamma_{M_I}^2}{4} \right) \left( \frac{1}{2} +
\frac{1}{\pi}\arctan{\left(\frac{\omega_{SR}+E_{M_I}-\omega_A}
{\Gamma_K/2}\right)} \right)\,,
\\
F_3&=& \frac{\Gamma_K}{2} \frac{\Gamma_{M_I}}{2}
\frac{\omega_N-\omega_A}{\pi^2} \ln{\left(
\frac{(\omega_{SR}+E_{M_I}-\omega_A)^2+(\Gamma_K/2)^2}
{(\omega_{SR}+E_{M_I}-\omega_N)^2+(\Gamma_{M_I}/2)^2}\right)}
\,.\nonumber
\end{eqnarray}
(We have neglected the nuclear level width because the following
relation is true $\Gamma_{N^*}\ll \Gamma_{M_I,K}$.) Formulae
(\ref{eq:sigma2})--(\ref{eq:F3}) give the cross section of nuclear
excitation in the NEET process near the $K$-shell photo-ionization
threshold for monochromatic photons. If the real SR beam has the
width $\Delta\omega_{SR}$, one should integrate the cross section
$\sigma_{N^*}$ over the beam line shape: $\int{}g(\omega-\omega_{SR})
\sigma_{N^*}(\omega)d\omega$. We used here the Gauss shape
$g(\omega-\omega_{SR}) = (1/\sqrt{\pi}\Delta\omega_{SR})
\exp{\left(-\left((\omega-\omega_{SR})/
\Delta\omega_{SR}\right)^2\right)}$. We will discuss some specific
examples in Sec.~\ref{sec:Results}.

Let us see now how the cross section $\sigma_{N^*}$ in
Eq~(\ref{eq:sigma1}) behaves in the range $\omega_{SR} \gg -E_K$,
i.e. far from the photo-ionization threshold. Let us consider once
more the expression for $f(E_e)$ in Eq~(\ref{eq:f}). The resonance
(``$\delta$-shaped'') nature of the function $f(E_e)$ is evident for
all values of $\omega_{SR}$. The function $f(E_e)$ is different from
zero in the range with the typical dimension $\Gamma_K+\Gamma_{M_I}$.
The remaining functions under the integration sign in Eq~(\ref{eq:sigma1})
change little in
the mentioned range of $E_e$ and can be taken outside the integral
sign. As a result we can consider just the limiting value of the
integral Eq~(\ref{eq:Int1}):

\begin{equation}
\lim_{\omega_{SR}\rightarrow\infty} \int_0^{\infty}f(E_e)dE_e =
\frac{1}{2\pi} \frac{\Gamma_K+\Gamma_{M_I}}
{(\omega_N-\omega_A)^2+(\Gamma_K+\Gamma_{M_I})^2/4} \,.
\label{eq:Lim}
\end{equation}
If we substitute this result to Eq~(\ref{eq:sigma2}) we obtain the
following relation for the cross section far from the threshold

\begin{equation}
\sigma_{N^*} = \sigma_{ion} P_{NEET}\,,
\label{eq:sigma3}
\end{equation}
where the relative probability $P_{NEET}$ was defined in
Eq~(\ref{eq:Pneet}). This result agrees well with an intuitive
conception about a factorization of the third order graph in
Fig.~\ref{fig:FD}(a). That process can be represented as a succession
of two processes in the incident photon energy range far from the
threshold (where the threshold effects are not significant). The
first process is an atomic shell photo-ionization, and the second
process is, in fact, the NEET. (It will be observed that a
contribution of the exchange diagram in Fig.~\ref{fig:FD}(b)
decreases as $1/\omega_{SR}^2$ at the range $\omega_{SR}\gg -E_K$,
i.e. this contribution is inessential as before.)

\section{Excitation of nuclei $^{197}$Au, $^{193}$Ir, and $^{189}$Os}
\label{sec:Results}

\begin{figure}[h]
\includegraphics[width=8cm]{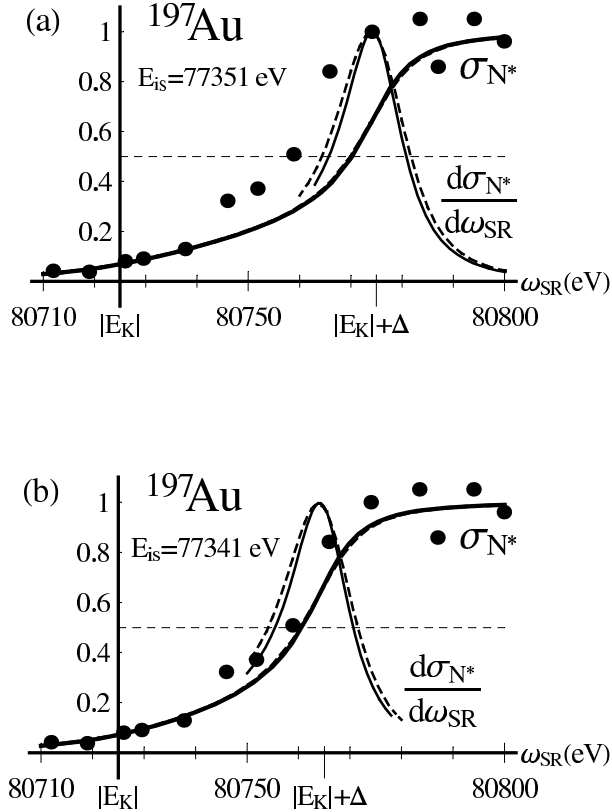}
\caption{The excitation cross section of the nucleus $^{197}$Au in
the units $\sigma_{ion}^0 P_{NEET}$ ($\simeq 1.3\times10^{-28}$
cm$^2$). Solid lines --- excitation by a monochromatic beam; dashed
lines --- excitation by the synchrotron radiation beam with 3.5 eV
width. Closed circles are the experimental data from
Ref.~\cite{Kishimoto-06}, Fig.3b. (a) The energy of the isomeric
level $E_{is}=77.351$ keV, and $\Delta=\omega_N-\omega_A=51$ eV; (b)
$E_{is}=77.341$ keV, and $\Delta=41$ eV.}
\label{fig:Au}
\end{figure}

\begin{figure}[h]
\includegraphics[width=8cm]{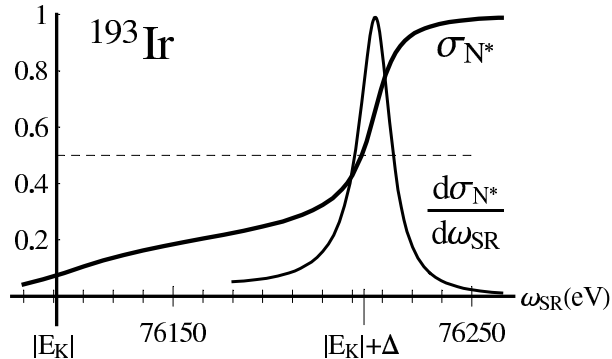}
\caption{The excitation cross section of $^{193}$Ir in the units
$\sigma_{ion}^0 P_{NEET}$($\simeq 6.6\times10^{-30}$ cm$^2$),
$\Delta=107$ eV. }
\label{fig:Ir}
\end{figure}

\begin{figure}[h]
\includegraphics[width=8cm]{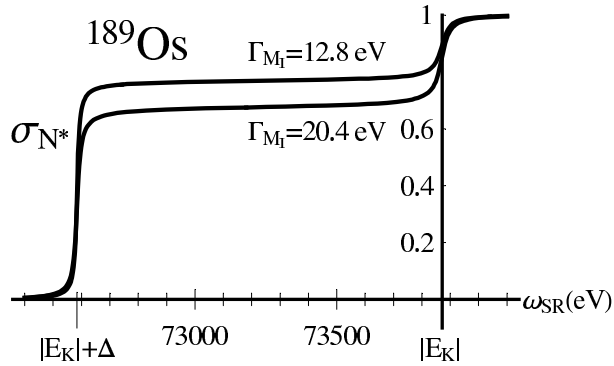}
\caption{Excitation of $^{189}$Os below the $K$-shell ionization
threshold for two $\Gamma_{M_I}$ width values. $\sigma_{N^*}$ is in
the units $\sigma_{ion}^0 P_{NEET}$ ($\simeq 4\times10^{-31}$
cm$^2$), $\Delta=-1287$ eV.}
\label{fig:Os}
\end{figure}

Here we consider how the cross section looks like for nuclei
$^{197}$Au, $^{193}$Ir, and $^{189}$Os near the $K$-shell ionization
threshold. The functions $\sigma_{N^*}(\omega_{SR})/\sigma_{ion}^0
P_{NEET}$ are shown in Figs.~\ref{fig:Au}--\ref{fig:Os} (i.e. all the
cross sections $\sigma_{N^*}$ in these Figures are in the units of
$\sigma_{ion}^0 P_{NEET}$). We denote else as $\Delta$ in
Figs.~\ref{fig:Au}--\ref{fig:Os} a difference between the energies of
nuclear transition and atomic transition, $\Delta \equiv
\omega_N-\omega_A$.

The following values of energies and widths were used for the cross
section $\sigma_{N^*}$ calculation at the nucleus $^{197}$Au by
formulae (\ref{eq:sigma2})--(\ref{eq:F3}): $E_K=-80.725$ keV,
$E_{M_I}=-3.425$ keV \cite{NIST}, $\Gamma_K=52$ eV
\cite{Campbell-01}, $\Gamma_{M_I}=14.3$ eV \cite{Campbell-01} (it
should be noted that there are four various values for
$\Gamma_{M_I}$: from 14.3 eV \cite{Campbell-01} to 20.9 eV
\cite{McGuire-72}, and here we use the value 14.3 eV recommended in
review \cite{Campbell-01}), $\omega_N=77.351$ keV \cite{Xiaolong-05}.
This nucleus is unique because the difference between the energies of
the atomic transition $\omega_A$ and the nuclear transition
$\omega_N$ is 50 eV only, i.e. this difference is commensurable with
the atomic widths in $K$ and $M_I$ shells of Au.

There are two kinds of lines in Fig.~\ref{fig:Au}. The solid lines
correspond to the excitation by a monochromatic beam. The dashed
lines correspond to the real SR-beam of the Gaussian form with the
width of 3.5 eV. It is evident from Fig.~\ref{fig:Au}(a) that the
nuclear excitation begins when the incident photon energy is above
the threshold by 49 eV approximately. The effective width of the
process (with the function $d\sigma_{N^*}/d\omega_{SR}$ (FWHM)) is
close to the $\Gamma_{M_I}$ width of Au for the monochromatic beam,
and it is equal to 17--18 eV for the real SR beam. These results are
in a qualitative agreement with the experimental data
\cite{Kishimoto-06}.


There are two relatively small differences between theoretical and
experimental data for the NEET process in $^{197}$Au at the present
time. Experimentalists found that the NEET events rose up above the
$K$ absorption edge and the NEET edge of a narrow width (= $14\pm9$
eV) existed at $40\pm2$ eV higher than the K edge
\cite{Kishimoto-06}. The theoretical values are approximately 19 eV
and 49 eV correspondingly. The NEET probability $P_{NEET}$ measured
in \cite{Kishimoto-06} was $(4.5\pm0.6)\times10^{-8}$, whereas the
theoretical value calculated by Eqs.\ (\ref{eq:Pneet}) and
(\ref{eq:Eint}) is $(3.4\pm 0.2)\times10^{-8}$ with the nuclear
reduced probability in Weisskopf units $B_{W.u.}(M1;1/2^+\rightarrow
3/2^+)=(2.05\pm 0.08)\times10^{-3}$ \cite{Xiaolong-05} and the atomic
matrix element evaluated here $|{\cal{}R}_1^M
(\omega_N;M_I\rightarrow{}K)|^2 =106.8$. It is interesting to note,
that the mentioned differences can be removed. The energy of the
isomeric level $E_{is}$ in $^{197}$Au is 77.351 keV
\cite{Xiaolong-05}. S.~Kishimoto suggested to consider another value
for the energy of the nuclear transition, viz 10 eV less than the
tabulated value \cite{Kishimoto-05-b}. In the case of using the value
$E_{is}=77.341$ keV in theoretical calculations, one obtains 40 eV
delay and 17.5 eV width for the NEET process near the $K$-edge (see
in Fig.~\ref{fig:Au} (b)), and $P_{NEET}=(4.5\pm 0.2)\times10^{-8}$
far from the $K$-edge. (We neglected here by the error bar for
$|{\cal{}R}_1^M (\omega_N;M_I\rightarrow{}K)|^2$, because the atomic
matrix elements are evaluated usually with the accuracy 0.1\% for the
transitions between the inner shells.) Such a good agreement with the
available experimental data points to the necessity to make a more
precise measurement of the energy of the first excited state in
$^{197}$Au. And in conclusion, the value $\sigma_{ion}^0
P_{NEET}\simeq 1.3\times10^{-28}$ cm$^2$ presented in
Fig.~\ref{fig:Au} was obtained with the experimental value
$P_{NEET}=4.5\times10^{-8}$ \cite{Kishimoto-06} and the tabulated
ionization cross section near the $K$ edge of Au $\sigma_{ion}^0 =
2.85\times10^{-23}$ cm$^2$ \cite{Veigele-73}.

The nuclear excitation cross section on $^{193}$Ir is shown in
Fig.~\ref{fig:Ir}. We used the following values of energies and
widths \cite{NIST,Campbell-01} for the calculation of the cross
section: $E_K=-76.111$ keV, $E_{M_I}=-3.174$ keV, $\Gamma_K=45$ eV,
$\Gamma_{M_I}=12.8$ eV, and $\omega_N=73.04$ keV
\cite{Artna-Cohen-98}. Calculation of $\sigma_{ion}^0 P_{NEET}$ gives
the following result $\sigma_{ion}^0 P_{NEET}\simeq
6.6\times10^{-30}$ cm$^2$. We used $\sigma_{ion}^0 =
3.28\times10^{-23}$ cm$^2$ from Ref.~\cite{Veigele-73} and $P_{NEET}
= 2.0\times10^{-9}$ evaluated with $B_{W.u.}(M1;1/2^+\rightarrow
3/2^+)=4.9\times10^{-4}$ \cite{Artna-Cohen-98}, and $|{\cal{}R}_1^M
(\omega_N;M_I\rightarrow{}K)|^2 =105.3$. The excitation of the
nucleus begins when the incident photon energy is above the threshold
--- $\omega_{SR}=-E_K+\omega_N-\omega_A$. The effective width of the
process is close to the $\Gamma_{M_I}$ width of Ir.


In $^{189}$Os the atomic transition energy
$\omega_A=E_{M_I}-E_K=70.822$ keV \cite{NIST} exceeds the energy of
the nuclear transition $\omega_N=69.535$ keV \cite{Wu-03} in contrast
to $^{197}$Au and $^{193}$Ir. Furthermore, the difference
$\omega_A-\omega_N=1287$ eV considerably exceeds the atomic width
values $\Gamma_K=42.6$ eV \cite{Campbell-01}, $\Gamma_{M_I}=20.4$ eV
\cite{McGuire-72}. The nuclear excitation cross section is shown in
Fig.~\ref{fig:Os}. We see that the excitation of the nucleus begins
when the incident photon energy is below the $K$-shell ionization
threshold: $\omega_{SR}=-E_K-1.287$ keV. The lower line in
Fig.~\ref{fig:Os} corresponds to the width $\Gamma_{M_I}=20.4$ eV
from Ref.~\cite{McGuire-72}, the upper line corresponds to the width
$\Gamma_{M_I}=12.8$ eV (this value is an approximation obtained from
the data in Ref.~\cite{Campbell-01}). The value of $\sigma_N$
averages 0.7--0.8 in the units $\sigma_{ion}^0 P_{NEET}$ in the range
$-E_K-(\omega_A-\omega_N) \leq{} \omega_{SR} \leq{}-E_K$. It means
that the SR photons effectively excite the nucleus below the
ionization threshold.

As regards the numerical value $\sigma_{ion}^0 P_{NEET}\simeq
4\times10^{-31}$ cm$^2$ in Fig.~\ref{fig:Os}, it was obtained with
$P_{NEET}=1.2\times10^{-10}$ \cite{Tkalya-92-NP} and the tabulated
ionization cross section near the $K$ edge of Os $\sigma_{ion}^0 =
3.4\times10^{-23}$ cm$^2$ \cite{Veigele-73}.


\begin{figure}[h]
\includegraphics[angle=-90,width=10cm]{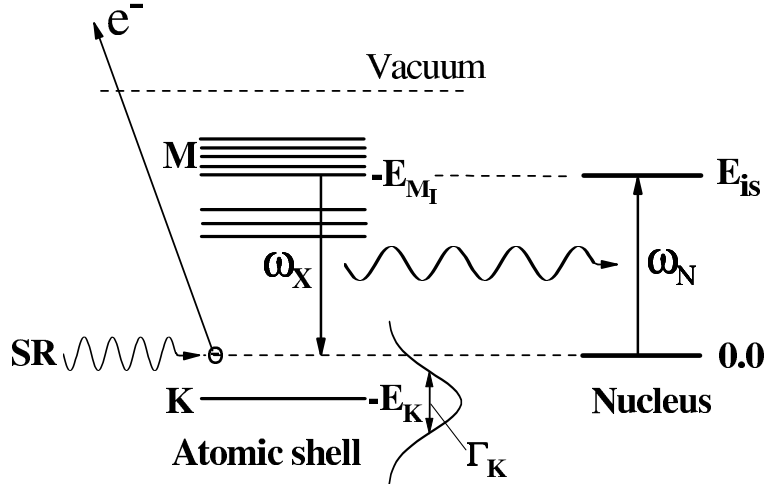}
\caption{Diagram of the subthreshold excitation of nucleus.}
\label{fig:Model}
\end{figure}

Subthreshold excitation is a quantum effect. The $K$-shell vacancy
has a sizeable width. As a consequence, incident photons with energy
below the binding energy $-E_K$ can ionize the $K$-shell. In this
case the energy of the emitted photon in the electron
$M_I\rightarrow{}K$ transition satisfies the condition
$\omega_X<E_{M_I}-E_K$ (see in Fig.~\ref{fig:Model}). On the other
hand the energy of the nuclear in $^{189}$Os transition satisfies the
analogous condition $\omega_N<\omega_A\equiv{}E_{M_I}-E_K$. That is
why the nuclear subthreshold excitation is possible. This very effect
is observed in Fig.~\ref{fig:Os}.


The $K$-shell hole is an intermediate state in such a process, and
the parameters of this state are not included into the energy
conservation law $(1/2\pi)\Gamma_{M_I}/
((\omega_{SR}-E_e+E_{M_I}-\omega_N)^2+\Gamma_{M_I}^2/4)$, which is an
analogue of the typical delta-function
$\delta(\omega_{SR}-E_e+E_{M_I}-\omega_N)$ connecting the initial and
final states in Fig.~\ref{fig:FD}. Therefore, the energy range where
there occurs the first (``basic'') resonance growth of the cross
section of the $^{189}$Os nucleus excitation (the area of the left
``step'' in Fig.~\ref{fig:Os}) is equal to $\Gamma_{M_I}$. For the
same reason at the interval with the width of $\Gamma_{M_I}$, but
beyond the ionization threshold, there also occurs the growth of the
$^{193}$Ir excitation. In $^{197}$Au the width of excitation cross
section is a bit greater than $\Gamma_{M_I}$. The difference between
the energies of atom and nucleus transitions in $^{197}$Au is about
the width of the vacancy for the $K$ shell of Au. As a result, the
second resonance at $\omega_{SR}= -E_K$ has an influence on the width
of excitation cross section at $\omega_{SR}=
-E_K+(\omega_N-\omega_A))$ and makes this basic resonance somewhat
broader. (We used the appellation ``basic'', because in the case of
wide interval between the resonances the amplitude of the resonance
at $\omega_{SR}= -E_K$ is smaller by a factor of
$\Gamma_K/\Gamma_{M_I}$ approximately, than the amplitude of the
basic resonance at $\omega_{SR}= -E_K+(\omega_N-\omega_A))$.

\section{Conclusion}

Finally, it should be noted that studying the processes of
interaction between a nucleus and an atom shell is quite a topical
task. One of the reasons is an active search, especially in recent
years, of acceleration mechanisms for isomeric state decay of nuclei
by an external influence upon an atom shell. A number of attempts of
this type were analyzed in paper \cite{Tkalya-05}. The NEET and other
similar processes become most important in this case. A perturbation
theory for the QED can best of all be used to systemize and to well
fit the mentioned phenomena, and the present work is another
confirmation of this view. The theory of the NEET developed within
the framework of perturbation theory for the QED gives results that
are in agreement with experimental data both qualitatively and
quantitatively in a wide range of energy including the range of
ionization threshold of atom.

\section{Acknowledgement}

I thank Prof. S.~Kishimoto for providing with the experimental data
and for a useful discussion on the NEET mechanism.


\begin{thebibliography}{23}
\expandafter\ifx\csname
natexlab\endcsname\relax\def\natexlab#1{#1}\fi
\expandafter\ifx\csname bibnamefont\endcsname\relax
  \def\bibnamefont#1{#1}\fi
\expandafter\ifx\csname bibfnamefont\endcsname\relax
  \def\bibfnamefont#1{#1}\fi
\expandafter\ifx\csname citenamefont\endcsname\relax
  \def\citenamefont#1{#1}\fi
\expandafter\ifx\csname url\endcsname\relax
  \def\url#1{\texttt{#1}}\fi
\expandafter\ifx\csname urlprefix\endcsname\relax\def\urlprefix{URL
}\fi \providecommand{\bibinfo}[2]{#2}
\providecommand{\eprint}[2][]{\url{#2}}

\bibitem[{\citenamefont{Morita}(1973)}]{Morita-73}
\bibinfo{author}{\bibfnamefont{M.}~\bibnamefont{Morita}},
  \bibinfo{journal}{Progr. Theor. Phys.} \textbf{\bibinfo{volume}{49}},
  \bibinfo{pages}{1574} (\bibinfo{year}{1973}).

\bibitem[{\citenamefont{Tkalya}(1992{\natexlab{a}})}]{Tkalya-92-NP}
\bibinfo{author}{\bibfnamefont{E.~V.} \bibnamefont{Tkalya}},
  \bibinfo{journal}{Nucl. Phys. A} \textbf{\bibinfo{volume}{539}},
  \bibinfo{pages}{209} (\bibinfo{year}{1992}{\natexlab{a}}).

\bibitem[{\citenamefont{Tkalya}(1992{\natexlab{b}})}]{Tkalya-92-JETP}
\bibinfo{author}{\bibfnamefont{E.~V.} \bibnamefont{Tkalya}},
  \bibinfo{journal}{Sov. Phys. JETP} \textbf{\bibinfo{volume}{75}},
  \bibinfo{pages}{200} (\bibinfo{year}{1992}{\natexlab{b}}),
  \bibinfo{note}{[Zh. Eksp. Teor. Fiz. {\bf 102}, 379 (1992)]}.

\bibitem[{\citenamefont{Tkalya}(2000)}]{Tkalya-00}
\bibinfo{author}{\bibfnamefont{E.~V.} \bibnamefont{Tkalya}},
  \bibinfo{journal}{AIP Conf. Proc.} \textbf{\bibinfo{volume}{506}},
  \bibinfo{pages}{486} (\bibinfo{year}{2000}).

\bibitem[{\citenamefont{Ahmad et~al.}(2000)\citenamefont{Ahmad, Dunford,
  Esbensen et~al.}}]{Ahmad-00}
\bibinfo{author}{\bibfnamefont{I.}~\bibnamefont{Ahmad}},
  \bibinfo{author}{\bibfnamefont{R.~W.} \bibnamefont{Dunford}},
  \bibinfo{author}{\bibfnamefont{H.}~\bibnamefont{Esbensen}},
  \bibnamefont{et~al.}, \bibinfo{journal}{Phys. Rev. C}
  \textbf{\bibinfo{volume}{61}}, \bibinfo{pages}{051304(R)}
  (\bibinfo{year}{2000}).

\bibitem[{\citenamefont{Harston}(2001)}]{Harston-01}
\bibinfo{author}{\bibfnamefont{M.~R.} \bibnamefont{Harston}},
  \bibinfo{journal}{Nucl. Phys. A} \textbf{\bibinfo{volume}{690}},
  \bibinfo{pages}{447} (\bibinfo{year}{2001}).

\bibitem[{\citenamefont{Tkalya}(1994)}]{Tkalya-94-JETP}
\bibinfo{author}{\bibfnamefont{E.~V.} \bibnamefont{Tkalya}},
  \bibinfo{journal}{JETP} \textbf{\bibinfo{volume}{78}}, \bibinfo{pages}{239}
  (\bibinfo{year}{1994}), \bibinfo{note}{[Zh. Eksp. Teor. Fiz. {\bf 105}, 449
  (1994)]}.

\bibitem[{\citenamefont{Kishimoto et~al.}(2000)\citenamefont{Kishimoto, Yoda,
  Seto et~al.}}]{Kishimoto-00}
\bibinfo{author}{\bibfnamefont{S.}~\bibnamefont{Kishimoto}},
  \bibinfo{author}{\bibfnamefont{Y.}~\bibnamefont{Yoda}},
  \bibinfo{author}{\bibfnamefont{M.}~\bibnamefont{Seto}}, \bibnamefont{et~al.},
  \bibinfo{journal}{Phys. Rev. Lett.} \textbf{\bibinfo{volume}{85}},
  \bibinfo{pages}{1831} (\bibinfo{year}{2000}).

\bibitem[{\citenamefont{Kishimoto et~al.}(2005)\citenamefont{Kishimoto, Yoda,
  Kobayashi et~al.}}]{Kishimoto-05}
\bibinfo{author}{\bibfnamefont{S.}~\bibnamefont{Kishimoto}},
  \bibinfo{author}{\bibfnamefont{Y.}~\bibnamefont{Yoda}},
  \bibinfo{author}{\bibfnamefont{Y.}~\bibnamefont{Kobayashi}},
  \bibnamefont{et~al.}, \bibinfo{journal}{Nucl. Phys. A}
  \textbf{\bibinfo{volume}{748}}, \bibinfo{pages}{3} (\bibinfo{year}{2005}).

\bibitem[{\citenamefont{Kishimoto et~al.}(2006)\citenamefont{Kishimoto, Yoda,
  Kobayashi et~al.}}]{Kishimoto-06}
\bibinfo{author}{\bibfnamefont{S.}~\bibnamefont{Kishimoto}},
  \bibinfo{author}{\bibfnamefont{Y.}~\bibnamefont{Yoda}},
  \bibinfo{author}{\bibfnamefont{Y.}~\bibnamefont{Kobayashi}},
  \bibnamefont{et~al.}, \bibinfo{journal}{Phys. Rev. C}
  \textbf{\bibinfo{volume}{74}}, \bibinfo{pages}{031301(R)}
  (\bibinfo{year}{2006}).

\bibitem[{\citenamefont{Berestetskii et~al.}(1982)\citenamefont{Berestetskii,
  Lifschitz, and Pitaevskii}}]{Berestetskii-80}
\bibinfo{author}{\bibfnamefont{V.~B.} \bibnamefont{Berestetskii}},
  \bibinfo{author}{\bibfnamefont{E.~M.} \bibnamefont{Lifschitz}},
  \bibnamefont{and} \bibinfo{author}{\bibfnamefont{L.~P.}
  \bibnamefont{Pitaevskii}}, \emph{\bibinfo{title}{Quantum Electrodynamics}}
  (\bibinfo{publisher}{Pergamon Press}, \bibinfo{address}{Oxford, England},
  \bibinfo{year}{1982}).

\bibitem[{\citenamefont{Bohr and Mottelson}(1969)}]{Bohr-69}
\bibinfo{author}{\bibfnamefont{A.}~\bibnamefont{Bohr}} \bibnamefont{and}
  \bibinfo{author}{\bibfnamefont{B.}~\bibnamefont{Mottelson}},
  \emph{\bibinfo{title}{Nuclear Structure. V.1. Single-Particle Motion}}
  (\bibinfo{publisher}{W.A. Benjamin Inc.}, \bibinfo{address}{New York},
  \bibinfo{year}{1969}).

\bibitem[{\citenamefont{Xiaolong and Chunmei}(2005)}]{Xiaolong-05}
\bibinfo{author}{\bibfnamefont{H.}~\bibnamefont{Xiaolong}} \bibnamefont{and}
  \bibinfo{author}{\bibfnamefont{Z.}~\bibnamefont{Chunmei}},
  \bibinfo{journal}{Nucl. Data Sheets} \textbf{\bibinfo{volume}{104}},
  \bibinfo{pages}{283} (\bibinfo{year}{2005}).

\bibitem[{\citenamefont{Artna-Cohen}(1998)}]{Artna-Cohen-98}
\bibinfo{author}{\bibfnamefont{A.}~\bibnamefont{Artna-Cohen}},
  \bibinfo{journal}{Nucl. Data Sheets} \textbf{\bibinfo{volume}{83}},
  \bibinfo{pages}{921} (\bibinfo{year}{1998}).

\bibitem[{\citenamefont{C.~Wu and Niu}(2003)}]{Wu-03}
\bibinfo{author}{\bibfnamefont{S.}~\bibnamefont{C.~Wu}} \bibnamefont{and}
  \bibinfo{author}{\bibfnamefont{H.}~\bibnamefont{Niu}},
  \bibinfo{journal}{Nucl. Data Sheets} \textbf{\bibinfo{volume}{100}},
  \bibinfo{pages}{1} (\bibinfo{year}{2003}).

\bibitem[{\citenamefont{Abramowitz and Stegun}(1969)}]{Abramowitz-64}
\bibinfo{author}{\bibfnamefont{M.}~\bibnamefont{Abramowitz}} \bibnamefont{and}
  \bibinfo{author}{\bibfnamefont{I.~A.} \bibnamefont{Stegun}},
  \emph{\bibinfo{title}{Handbook of Mathematical Functions}}
  (\bibinfo{publisher}{National Bureau of Standards},
  \bibinfo{address}{Washington, D.C.}, \bibinfo{year}{1969}).

\bibitem[{\citenamefont{Band and Fomichev}(1979)}]{Band-79}
\bibinfo{author}{\bibfnamefont{I.~M.} \bibnamefont{Band}} \bibnamefont{and}
  \bibinfo{author}{\bibfnamefont{V.~I.} \bibnamefont{Fomichev}},
  \bibinfo{journal}{At. Data Nucl. Data Tabl.} \textbf{\bibinfo{volume}{23}},
  \bibinfo{pages}{295} (\bibinfo{year}{1979}).

\bibitem[{NIS()}]{NIST}
\bibinfo{note}{NIST Physical Reference Data},
  \eprint{http://physics.nist.gov/PhysRefData/contents.html}.

\bibitem[{\citenamefont{Campbell and Papp}(2001)}]{Campbell-01}
\bibinfo{author}{\bibfnamefont{J.~L.} \bibnamefont{Campbell}} \bibnamefont{and}
  \bibinfo{author}{\bibfnamefont{T.}~\bibnamefont{Papp}}, \bibinfo{journal}{At.
  Data Nucl. Data Tabl.} \textbf{\bibinfo{volume}{77}}, \bibinfo{pages}{1}
  (\bibinfo{year}{2001}).

\bibitem[{\citenamefont{McGuire}(1972)}]{McGuire-72}
\bibinfo{author}{\bibfnamefont{E.~J.} \bibnamefont{McGuire}},
  \bibinfo{journal}{Phys. Rev. A} \textbf{\bibinfo{volume}{5}},
  \bibinfo{pages}{1043} (\bibinfo{year}{1972}).

\bibitem[{Kis()}]{Kishimoto-05-b}
\bibinfo{note}{S. Kishimoto, Private Communication, October 2005}.

\bibitem[{\citenamefont{Veigele}(1973)}]{Veigele-73}
\bibinfo{author}{\bibfnamefont{W.}~\bibnamefont{Veigele}},
  \bibinfo{journal}{At. Data Tabl.} \textbf{\bibinfo{volume}{5}},
  \bibinfo{pages}{1} (\bibinfo{year}{1973}).

\bibitem[{\citenamefont{Tkalya}(2005)}]{Tkalya-05}
\bibinfo{author}{\bibfnamefont{E.~V.} \bibnamefont{Tkalya}},
  \bibinfo{journal}{Physics Uspekhi} \textbf{\bibinfo{volume}{48}},
  \bibinfo{pages}{525} (\bibinfo{year}{2005}), \bibinfo{note}{[Uspekhi Fiz.
  Nauk {\bf 175}, 555 (2005)]}.

\end{thebibliography}

\end{document}